\documentclass[twocolumn,preprintnumbers,amsmath,amssymb,prl]{revtex4}

\usepackage{graphicx}
\usepackage{dcolumn}
\usepackage{bm}
\usepackage{color}

\begin{document}

\title{
Tuning the pairing interaction in a $d$-wave superconductor \\by paramagnons injected through interfaces
}

\author{M.\,Naritsuka$^{1}$}
\author{P.F.S.\,Rosa$^{2}$}
\author{Y.\,Luo$^{2}$}
\author{Y.\,Kasahara$^1$}
\author{Y.\,Tokiwa$^{1,3}$}
\author{T.\,Ishii$^1$}
\author{S.\,Miyake$^1$}
\author{T.\,Terashima$^1$}
\author{T.\,Shibauchi$^4$}
\author{F.\,Ronning$^2$}
\author{J.D.\,Thompson$^2$}
\author{Y.\,Matsuda$^1$}

\affiliation{$^1$Department of Physics, Kyoto University, Kyoto 606-8502 Japan}
\affiliation{$^2$Los Alamos National Laboratory, Los Alamos, NM 87544, USA }
\affiliation{$^3$Center for Electronic Correlations and Magnetism, Institute of Physics, Augsburg University, 86159 Augsburg, Germany} 
\affiliation{$^4$Department of Advanced Materials Science, University of Tokyo, Chiba 277-8561, Japan} 


                             
\begin{abstract}

Unconventional superconductivity and magnetism are intertwined on a microscopic level in a wide class of materials.  A new approach to this most fundamental and hotly debated issue focuses on the role of interactions between superconducting electrons and bosonic fluctuations at the interface between adjacent layers in heterostructures.  
Here we fabricate hybrid superlattices consisting of alternating atomic layers of heavy-fermion superconductor CeCoIn$_5$ and antiferromagnetic (AFM) metal CeRhIn$_5$, in which the AFM order can be suppressed by applying pressure. 
We find that the superconducting and AFM states coexist in spatially separated layers, but their mutual coupling via the interface significantly modifies the superconducting properties. An analysis of upper critical fields reveals that near the critical pressure where AFM order vanishes, the force binding superconducting electron-pairs acquires an extremely strong-coupling nature. This demonstrates that superconducting pairing can be tuned non-trivially by magnetic fluctuations (paramagnons) injected through the interface, leading to maximization of the pairing interaction.

\end{abstract}

\maketitle


In diverse families of strongly correlated electron systems, including cuprates,  iron-pnictides,  and heavy fermion compounds, superconductivity is often found near a quantum critical point (QCP) where a magnetic phase vanishes in the limit of zero temperature, pointing to a magnetic glue as the source of electron pairing \cite{Mathur,Shibauchi2014,Keimer}.   In these materials,  microscopic coexistence of superconducting and magnetically ordered phases both involving the same charge carriers is a striking example for unusual emergent electronic phases.  Moreover,  superconductivity is frequently the strongest near the QCP,  suggesting that the proliferation of critical magnetic excitations emanating from the QCP plays an important role in Cooper pairing.   Despite tremendous research, however,  the entangled relationship between superconductivity and magnetism has remained largely elusive.

\begin{figure}[b]
\begin{center}
\includegraphics[width=1.0\linewidth]{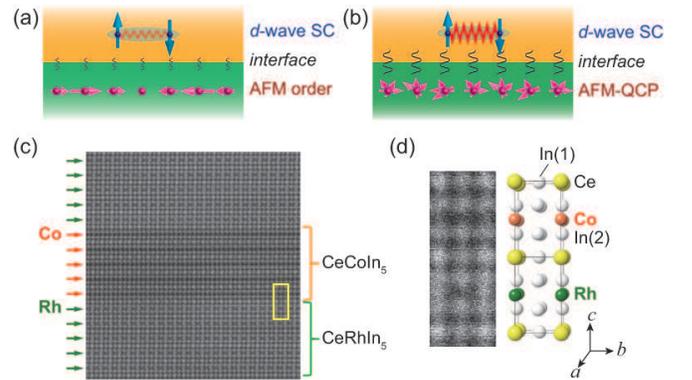}
\caption{
(a) Schematic figure of the interaction between $d$-wave superconductivity (SC) and static antiferromagnetic (AFM) order  via the interface.  (b) Interaction between two competing orders under pressure near quantum critical point (QCP), where  AFM order disappears. 
(c) High resolution cross-sectional TEM image for CeCoIn$_5$($5$)/CeRhIn$_5$($5$) superlattice. (d) The TEM image of the boxed area in (c). 
}
 \end{center}
 \end{figure}

Recently, realization that interactions between superconducting electrons and bosonic excitations through an atomic interface can have a profound influence on Cooper-pair formation has raised the exciting possibility of a new route to controlling superconductivity.   For instance, when a monolayer of FeSe is grown on a SrTiO$_3$ substrate, the interaction between FeSe electrons and SrTiO$_3$ phonons via the interface enhances the pairing interaction, giving rise to the highest  transition temperature $T_c$ among iron-based superconductors \cite{Huang,DHLee,JJLee,Rademaker}.    This discovery raises the possibility of a magnetic analogue in which the pairing interaction is influenced by magnetic fluctuations though an interface between  an unconventional superconductor and a magnetic metal. This concept is illustrated schematically in Figs.\,1(a) and 1(b). Besides allowing a new approach to revealing the entangled relationship between magnetism and unconventional superconductivity, this concept has the  advantage that magnetic excitations are tunable as a magnetic transition is driven toward zero temperature, unlike phonon excitations in SrTiO$_3$.  The state-of-the-art molecular beam epitaxy (MBE) technique enables realization of this idea through fabrication of  artificial Kondo superlattices with alternating layers of Ce-based heavy fermion superconductors and magnets  that are atomic layer thick \cite{Shishido2010,Mizukami,Shimozawa2016}.  These artificially engineered materials are particularly suitable systems to elucidate the mutual interaction through the interface, providing a new platform to study the interplay of competing orders.  
 
The layered heavy fermion compounds Ce$M$In$_5$ ($M=$\,Co, Rh) are ideal model systems in which the interplay between magnetism and superconductivity can be explored, because of their high purity and small energy scales \cite{Thompson,Kenzelman,Knebel2010}.  They  have similar Fermi surface structures and similar pressure-temperature ($p$-$T$) phase diagrams.    At ambient pressure, CeCoIn$_5$ is a superconductor ($T_c$=2.3\,K) with $d_{x^2-y^2}$-wave symmetry \cite{Izawa,Allan,Zhou}. The normal state possesses non-Fermi-liquid properties in zero field, including $T$-linear resistivity, indicative of a nearby underlying QCP \cite{Sidorov,Nakajima}.  In contrast, CeRhIn$_5$ orders antiferromagnetically at atmospheric pressure ($T_{\rm N}$=3.8\,K) \cite{Bao}.  Its magnetic transition is suppressed by applying pressure and the ground state becomes purely superconducting state at $p>p^\ast\approx$1.7\, GPa, indicating the presence of a pressure induced QCP \cite{Park2006,Knebel2008,Park2008,Shishido2005}.   As disorder may seriously influence physical properties especially near a QCP, there is a great benefit in examining quantum critical systems which are stoichiometric, and hence, relatively disorder free; both compounds are ones of a small number of such systems. Both host a wide range of fascinating superconducting properties including an upper critical field $H_{c2}$ that is limited by  extremely strong Pauli pair-breaking \cite{Izawa,Knebel2008}. 

To realize hybrid heterostructures shown in Figs.\,1(a) and 1(b), we fabricate superlattice films with alternating block layers (BLs) of $n$ unit-cell-thick (UCT) CeCoIn$_5$ and $m$-UCT CeRhIn$_5$, CeCoIn$_5$($n$)/CeRhIn$_5$($m$). 
We demonstrate that the pairing interaction in a $d$-wave superconductor is tuned by injecting magnetic fluctuations through the atomic interface. Moreover, we show that the pairing strength is maximized near the critical pressure where AFM order vanishes.


The hybrid superlattices CeCoIn$_5$($n$)/CeRhIn$_5$($m$) with $c$ axis oriented structure are grown on MgF$_2$ substrate by the MBE technique \cite{Shishido2010,Mizukami,Shimozawa2016}. 
Figure\,1(c) displays a high-resolution cross-sectional transmission electron microscope (TEM) image of a CeCoIn$_5$($5$)/CeRhIn$_5$($5$) superlattice.  The TEM image displayed in Fig.\,1(d) (the boxed area in Fig.\,1(c)) demonstrate that the Rh and Co atoms are clearly distinguished by bright and dark spots, respectively.  No discernible atomic inter-diffusion between the neighboring Co and Rh layers is seen, which is also confirmed by lateral satellite peaks in an X-ray diffraction pattern.  The epitaxial growth of each layer with atomic flatness is confirmed by  reflection high energy electron diffraction (Fig.\,S1 in \cite{SM}). These results indicate the successful fabrication of epitaxial superlattices with sharp interfaces.  
High-pressure resistivity measurements have been performed under hydrostatic pressure up to 2.4\,GPa using a piston cylinder cell with oil as pressure transmitting medium. 


 \begin{figure}[t]
 	\begin{center}
 		\includegraphics[width=1.0\linewidth]{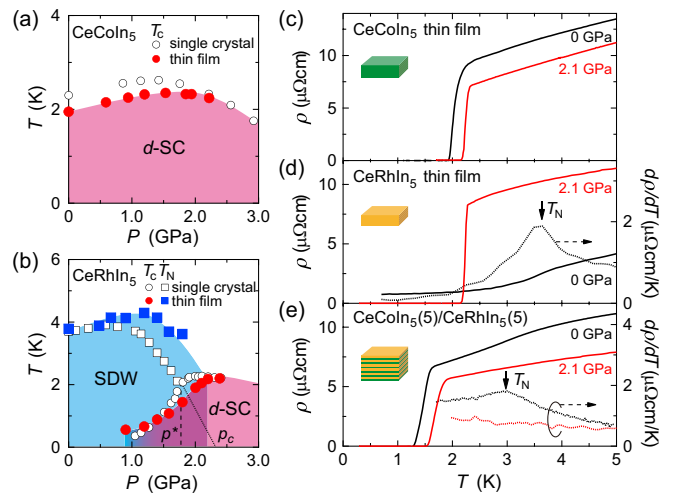}
 		\caption{
		(a), (b) $p$-$T$ phase diagrams of thin films and single crystals of (a) CeCoIn$_5$  and (b) CeRhIn$_5$. 
		(c) Temperature dependence of the resistivity of CeCoIn$_5$ thin film at ambient pressure and at $p=2.1$\,GPa. (d) and (e) show temperature dependence of the resistivity (solid lines, left axes) and its temperature derivative $d\rho(T)/dT$ (dotted lines, right axes) for CeRhIn$_5$ thin film and CeCoIn$_5$(5)/CeRhIn$_5$(5) superlattice at ambient pressure and at $p=2.1$\,GPa, respectively. The peak of $d\rho(T)/dT$ corresponds to AFM transition.  
 		}
 		\label{fig:Fig1.eps}
 	\end{center}
 \end{figure}

Figures\,2(a) and 2(b) depict the  resistively determined  $p$-$T$ phase diagrams of  separate, MBE-grown epitaxial thin films of  CeCoIn$_5$ and CeRhIn$_5$, whose resistivities  ($\rho$) are  shown in Figs.\,2(c) and 2(d), respectively.  The $p$-$T$ phase diagrams of both films are essentially those of single crystals.  
$T_c$ (=2.0\,K) in the CeCoIn$_5$ thin film, however, is slightly reduced from the bulk value, possibly due to strain  induced by a slight lattice mismatch with  the substrate, while $T_{\rm N}$ (=3.7\,K) of CeRhIn$_5$ thin film is almost the same as that in a single crystal.  With pressure, $T_c$ of CeCoIn$_5$ thin film increases and shows a broad peak  near $p\sim$1.7\,GPa.  CeRhIn$_5$ thin film undergoes the superconducting transition with no signature of AFM transition at $p\approx$2.1\,GPa.  Similar to CeRhIn$_5$  single crystals \cite{Park2006,Sidorov,Park2008},  superconductivity  in the thin films develops  at $p\agt$1\,GPa  where it coexists with magnetic order, and  there is only a  purely superconducting state  at $p\agt$2.1\,GPa (Fig.\,S2 in \cite{SM}), a slightly higher pressure than in single crystals.

\begin{figure}[t]
 	\begin{center}
 		\includegraphics[width=1.0\linewidth]{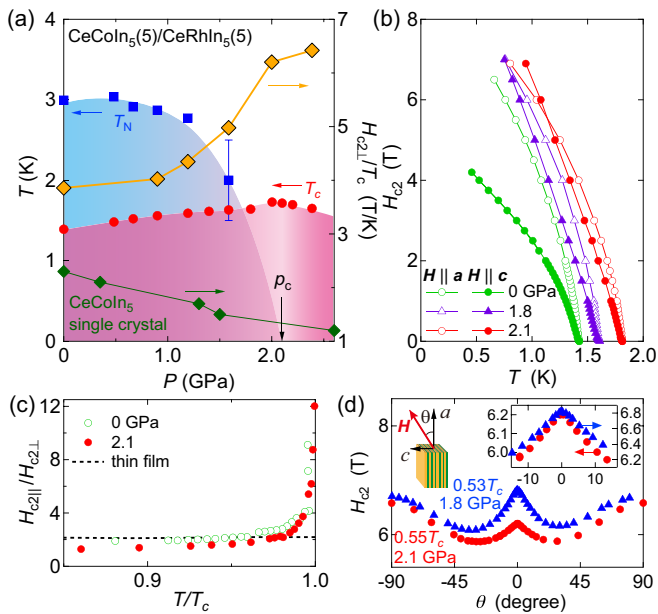}
 		\caption{
		(a) $p$-$T$ phase diagram of CeCoIn$_5$(5)/CeRhIn$_5$(5) superlattice.  
		Out-of-plane upper critical field $H_{c2\perp}$ normalized by $T_c$, $H_{c2\perp}/T_c$, measures the coupling strength of the  superconductivity. 
		(b) Temperature dependence of in-plane and out-of-plane upper critical fields at ambient pressure and at $p=1.8$ and 2.1\,GPa. (c) Anisotropy of upper critical field, $H_{c2\parallel}/H_{c2\perp}$, near $T_c$ of superlattices at ambient pressure and at 2.1\,GPa, along with the data of CeCoIn$_5$ thin film. (d) Angular dependence of upper critical field of superlattice at $p=1.8$ and 2.1\,GPa. The inset is an expanded view of the low angle region.
 		}
 		\label{fig:Fig1.eps}
 	\end{center}
 \end{figure}

Figure\,2(e)  compares  the $T$-dependence of $\rho(T)$ and its temperature derivative $d\rho(T)/dT$ for a CeCoIn$_5$(5)/CeRhIn$_5$(5) superlattice at ambient pressure and at $p=2.1$\,GPa.   At ambient pressure, a distinct peak in $d\rho(T)/dT$ associated with an AFM transition can be seen at 3\,K in addition to a superconducting transition at $\sim1.4$\,K \cite{Knebel2008}. While $T_c$ and $T_{\rm N}$ of the hybrid superlattice are lower than that of the CeCoIn$_5$ and CeRhIn$_5$ thin films, respectively, they are still larger than that of respective CeCoIn$_5$/YbCoIn$_5$ and CeRhIn$_5$/YbRhIn$_5$ superlattices (Fig.\,S2 in \cite{SM}), indicating the importance of mutual interaction between the CeCoIn$_5$ and CeRhIn$_5$ BLs.  On the other hand, at $p=2.1$\,GPa, there is no signature for magnetic order, while the superconductivity remains with slightly higher $T_c$ than at ambient pressure.  In Fig.\,3(a),  we plot the $p$-dependence of $T_c$ and $T_{\rm N}$ determined by the peak in $d\rho(T)/dT$.   At $p\sim2$\,GPa,  $T_c$ is a maximum, forming a dome-shaped $p$-dependence.  With pressure, $T_{\rm N}$ is suppressed gradually at low $p$, followed by a rapid suppression at $p\agt1$\,GPa (Fig.\,S3 in \cite{SM}).   At $p\agt 1.6$\,GPa, evidence for magnetic order is hidden beneath the superconducting dome. A simple extrapolation of $T_{\rm N}(p)$ gives a critical pressure $p_c\sim2$\,GPa at which the magnetic transition reaches zero temperature and $T_c$ shows a maximum. 

We demonstrate that two-dimensional (2D) superconductivity is realized  in CeCoIn$_5$ BLs in the whole pressure regime.   Figures\,3(b) and 3(c) depict the $T$-dependence of the upper critical field determined by the mid point of the resistive transition in a magnetic field $H$ applied  parallel  ($H_{c2\parallel}$)  and perpendicular  ($H_{c2\perp}$)  to the $ab$ plane and the $T$-dependence of the anisotropy of upper critical fields, $H_{c2\parallel}/H_{c2\perp}$, respectively.  The anisotropy diverges on approaching $T_c$, in sharp contrast to the CeCoIn$_5$ thin film whose anisotropy shows little $T$-dependence up to $T_c$.  This diverging anisotropy in the superlattice is a characteristic feature of 2D superconductivity, in which  $H_{c2\parallel}$ increases as $\sqrt{T_c-T}$ due to the Pauli paramagnetic limiting, but   $H_{c2\perp}$ increases as $T_c-T$ due to orbital limiting near $T_c$.    This result, along with the fact that the thickness of the CeCoIn$_5$-BL is comparable to the perpendicular superconducting coherence length $\xi_{\perp}\sim3$--4\,nm, indicates that each 5-UCT CeCoIn$_5$ BL effectively acts as a 2D superconductor \cite{Mizukami}.   The 2D superconductivity is reinforced by the angular variation of $H_{c2}(\theta)$.  Figure \,3(d) and its inset show $H_{c2}(\theta)$ below and above $p^*$.   For both pressures, at $T\ll T_c$, 
$H_{c2}(\theta)$ in the regime $|\theta|\alt30^{\circ}$ is enhanced with decreasing $|\theta|$ and exhibits a sharp cusp at $\theta=0$.  This cusp behavior is typical for a Josephson coupled layered superconductor \cite{Tinkham}.

We note that  in stark contrast to CeRhIn$_5$ single crystal and our thin film, each CeRhIn$_5$ BL in CeCoIn$_5$(5)/CeRhIn$_5$(5) superlattice is not fully superconducting even when the AFM order is suppressed under pressure, which leads to the realization of 2D superconductivity in a wide range of pressure.  In fact, as shown in Fig.\,3(d), overall angle dependence of $H_{c2}(\theta)$ including the cusp structure near $\theta=0$ is observed at $p=1.8$\,GPa, where the bulk superconductivity is not observed in CeRhIn$_5$ thin film (Fig.\,2(b) and Fig.\,S2 in \cite{SM}).  Essentially very similar angle dependence of $H_{c2}(\theta)$ is observed at $p=2.1$\,GPa above $p_c$.  These results imply that 2D superconductivity occurs in CeCoIn$_5$ BLs even above $p_c$.   Moreover,  in CeRhIn$_5$(5)/YbRhIn$_5$(5) 
superlattice zero resistivity is not attained under pressure (Fig.\,S4 in \cite{SM}).   
With the reduction of BL thickness, the superconductivity of CeRhIn$_5$ is strongly suppressed, in stark contrast to CeCoIn$_5$. This may be related to the incommensurate magnetic structure of CeRhIn$_5$ with ordering vector $\bm{q}=(0.5,0.5, 0.297)$ \cite{Bao}, in which the long-wave-length AFM fluctuations perpendicular to the layers are suppressed in CeRhIn$_5$ BLs with atomic layer thickness.  In CeCoIn$_5$, on the other hand, AFM fluctuations with different $\bm{q}=(0.45, 0.45, 0.5)$ are dominant \cite{Raymond}. This commensurability along the $c$ axis would be better compatible with the superlattice structure, and as a result,   the superconductivity is robust against the reduction of BL thickness \cite{Yamanaka}. 
We here comment on the low temperature anisotropy of $H_{c2}$ of the CeCoIn$_5$(5)/CeRhIn$_5$(5) superlattice  (Fig.\,3(b)).  At $p=2.1$\,GPa, $H_{c2\perp}$ exceeds $H_{c2\parallel}$ at low temperatures. Such a  reversed  anisotropy of $H_{c2}$ has been reported in CeRhIn$_5$ single crystal above the pressure where the AFM order disappears \cite{Thompson,Park2008}.  However, similar reversed anisotropy ($H_{c2\perp}>H_{c2\parallel}$) is preserved at $p=1.8$\,GPa, where $H_{c2\parallel}$ exceeds $H_{c2\perp}$ in CeRhIn$_5$ single crystal and thin film. This indicates that anisotropy reversal of $H_{c2}$ occurs under pressure in  5-UCT CeCoIn$_5$ BLs.  Based on these results, we conclude that  2D superconducting CeCoIn$_5$ BLs in CeCoIn$_5$(5)/CeRhIn$_5$(5) are coupled by the Josephson effect in the whole pressure regime.

\begin{figure}[t]
 	\begin{center}
 		\includegraphics[width=1.0\linewidth]{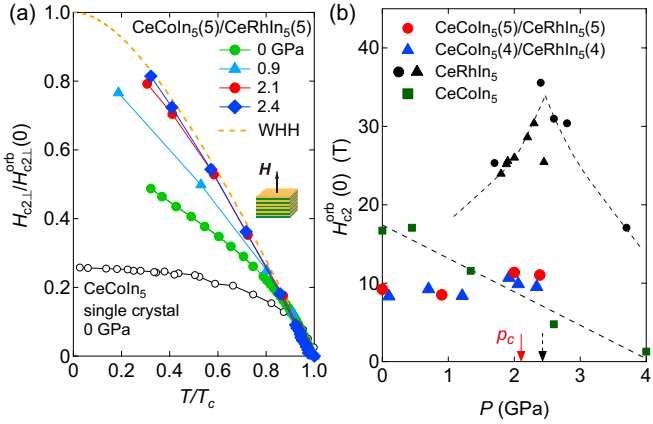}
 		\caption{
		(a) Out-of-plane upper critical field $H_{c2\perp}$ normalized by the orbital-limited upper critical field at $T=0$\,K, $H_{c2\perp}/H_{c2\perp}^{\rm orb}(0)$, for CeCoIn$_5$(5)/CeRhIn$_5$(5) superlattice is plotted as a function of the normalized temperature $T/T_c$.   Two extreme cases, i.e.  the result of  the bulk CeCoIn$_5$ dominated by Pauli paramagnetic effect and the WHH curve with no Pauli effect, are also shown.  (b) Pressure dependence of $H_{c2}^{\rm orb}(0)$ of CeCoIn$_5$($n$)/CeRhIn$_5$($n$) superlattices with $n=4$ and 5 for ${\bm H}$$\parallel$$c$.   For comparison, $H_{c2}^{\rm orb}(0)$ of CeRhIn$_5$ single crystals  for ${\bm H}$$\parallel$$a$ and that of CeCoIn$_5$ single crystal for ${\bm H}$$\parallel$$c$ are shown. 
Solid and dashed arrows represent $p_c$ for CeCoIn$_5$($n$)/CeRhIn$_5$($n$) superlattices and CeRhIn$_5$ single crystal, respectively. 
 		}
 		\label{fig:Fig1.eps}
 	\end{center}
 \end{figure}


Application of the pressure leads to a drastic change in the nature of  superconductivity in the hybrid superlattices.   Figure\,4(a) depicts the $T$-dependence of $H_{c2\perp}$, normalized by the orbital-limited upper critical field at $T=0$\,K, $H_{c2\perp}^{\rm orb}(0)$, which is obtained from  the Werthamer-Helfand-Hohenberg (WHH) formula, $H_{c2\perp}^{\rm orb}(0)=-0.69T_c(dH_{c2\perp}/dT)_{T_c}$.  We also include two extreme cases: $H_{c2\perp}/H_{c2\perp}^{\rm orb}(0)$ for  bulk CeCoIn$_5$ \cite{Tayama}, in which $H_{c2}$  is dominated by Pauli paramagnetism, and the WHH curve with no Pauli effect.   Pressure dramatically enhances $H_{c2\perp}/H_{c2\perp}^{\rm orb}$.  What is remarkable is that  near  the critical pressure $p_c\sim 2$\,GPa at which  evidence for magnetic order  disappears, $H_{c2\perp}/H_{c2\perp}^{\rm orb}$ nearly coincides with the WHH curve, indicating that $H_{c2\perp}$ is limited solely by orbital pair-breaking.

The fact that  $H_{c2\perp}$ approaches the orbital limit provides important insight on superconductivity of the hybrid superlattice. In CeCoIn$_5$/YbCoIn$_5$, where YbCoIn$_5$ is a conventional metal, Pauli pair-breaking effect is weaken in the superlattice compared with the bulk due to local inversion symmetry breaking at the interfaces, which splits the Fermi surfaces with spin texture and thus effectively suppresses the Zeeman effect \cite{Goh,Maruyama2012}. This leads to the Rashba-induced anisotropic suppression of the Zeeman effect \cite{Shimozawa2016}, which may be partly responsible for the observed reversed anisotropy $H_{c2\parallel}/H_{c2\perp}<1$ at low temperatures (Fig.\,3(d)). However, this effect is less important in CeCoIn$_5$($n$)/CeRhIn$_5$($n$) superlattices compared with  CeCoIn$_5$/YbCoIn$_5$, which is evidenced by the fact that  $H_{c2\perp}/H_{c2\perp}^{\rm orb}(0)$ does not strongly depend on $n$ (Fig.\,S5 in \cite{SM}). Moreover, such an effect is not expected to have significant pressure dependence. Therefore, there must be a different mechanism that significantly enhances the Pauli-limiting field $H_{c2\perp}^{\rm Pauli}=\sqrt{2}\Delta/g\mu_B$,  where $g$ is the $g$-factor of electrons and $\mu_B$ is the Bohr magneton.  An enhancement of $H_{c2\perp}^{\rm Pauli}$ is not due to a dramatic suppression of $g$ by pressure, because it is highly unlikely that the Ce crystalline electric field state, which determines $g$-factor, strongly depends on pressure.  Therefore the enhancement of  $H_{c2\perp}^{\rm Pauli}$ is attributed to a strong increase in the superconducting gap $\Delta$.  This is supported by the observed enhancement of  $H_{c2\perp}/T_c$ upon approaching $p_c$ shown in Fig.\,3(a).    Because  $H_{c2\perp}\approx H_{c2\perp}^{\rm Pauli} \ll H_{c2\perp}^{\rm orb}(0)$ in the low $p$ regime and  $H_{c2\perp}\approx H_{c2\perp}^{\rm orb}(0) \ll H_{c2\perp}^{\rm Pauli}$ near $p\sim p_c$,   the enhancement of  $H_{c2\perp}/k_BT_c$ directly indicates an enhancement of $H_{c2\perp}^{\rm Pauli}/T_c$ and hence $\Delta/k_BT_c$.  This behavior  contrasts with observations on  CeCoIn$_5$ single crystals, in which $H_{c2}/T_c$ decreases with pressure.  The enhancement of $\Delta/k_BT_c$ is caused as a consequence of enhancement of pairing interaction. In spin fluctuation mediated scenario, the pairing interaction is mainly provided by high energy spin fluctuations whose energy scale is well above $\Delta$ and  low energy fluctuations cause the pair-breaking.  Since the high energy fluctuations enhance $T_c$ while low energy ones reduce $T_c$,  the enhancement of pairing interaction can give rise to the increase of $\Delta/k_BT_c$ without accompanying a large enhancement of $T_c$, which is consistent with the observed behavior.  Thus, the present results demonstrate that  the pairing interaction in CeCoIn$_5$ BLs is strikingly enhanced as a result of the quantum critical magnetic fluctuations  that develop in  CeRhIn$_5$ BLs, which are injected into CeCoIn$_5$ BLs  through the interface.

It is well established that quantum fluctuations strongly influence normal and superconducting properties in many classes of unconventional superconductors.   One of the most striking is a diverging effective quasiparticle mass $m^*$ upon approaching the QCP, as reported in cuprate, pnictide and heavy-fermion systems \cite{Shibauchi2014,Shishido2005,Ramshaw}.  Such a mass enhancement gives rise to a corresponding enhancement of $H_{c2}^{\rm orb}$, which is proportional to  $(m^*\Delta)^2$.  Here we stress that there is a fundamental difference in the present hybrid superlattices.  Figure\,4(b) depicts the $p$-dependence of $H_{c2\perp}^{\rm orb}$ of the CeCoIn$_5$($n$)/CeRhIn$_5$($n$) superlattices with $n=4$ and 5, along with the result for CeCoIn$_5$ and CeRhIn$_5$ single crystals \cite{Park2008,Knebel2010}.  In contrast to a CeRhIn$_5$ single crystal which shows a sharp peak at the critical pressure, $H_{c2\perp}^{\rm orb}$ of the superlattices depends weakly on pressure with no significant anomaly at $p_c$. Compared to the monotonic decrease  observed in single crystal CeCoIn$_5$, this weak dependence is consistent with an enlarged gap $\Delta$, but the results suggest the absence of 
mass enhancement in the CeCoIn$_5$ BL. Such a behavior is in contrast to usual expectations for quantum criticality, details of which deserve further studies.


In summary, we have designed and fabricated hybrid superlattice CeCoIn$_5$/CeRhIn$_5$ formed by alternating atomically thick layers of a $d$-wave heavy fermion superconductor CeCoIn$_5$ and an AFM metal CeRhIn$_5$. 
The present results demonstrate the importance of the interface between which unconventional superconducting and nonsuperconducting magnetic layers can interact with each other.  
In particular, the  strength of the pairing interaction can be tuned by magnetic fluctuations, or paramagnons, injected through the interface, highlighting that the pairing interaction can be maximized by the critical fluctuations emanating from the magnetic QCP without an accompanying mass enhancement.     The fabrication of a wide variety of hybrid superlattices  paves a new way to study the entangled relationship between unconventional superconductivity and magnetism, offering a route to exploring the emergence of novel superconducting systems and the roles of their interface.

We thank E.-A. Kim, H. Kontani, A. H. Nevidomskyy, R. Peters, and Y. Yanase for fruitful discussions. This work was supported by Grants-in-Aid for Scientific Research (KAKENHI) (Nos. 25220710, 15H02014, 15H02106, and 15H05457) and on Innovative Areas `Topological Material Science' (No. JP15H05852) and `3D Active-Site Science' (No. 26105004) from Japan Society for the Promotion of Science (JPSJ). Work at Los Alamos National Laboratory was performed under the auspices of the U.S. Department of Energy, Office of Basic Energy Sciences, Division of Materials Sciences and Engineering.


\clearpage

\renewcommand{\thefigure}{S\arabic{figure}}
\setcounter{figure}{0}

\section*{Supplemental Material}

\begin{figure}[h]
	\includegraphics[width=0.9\linewidth]{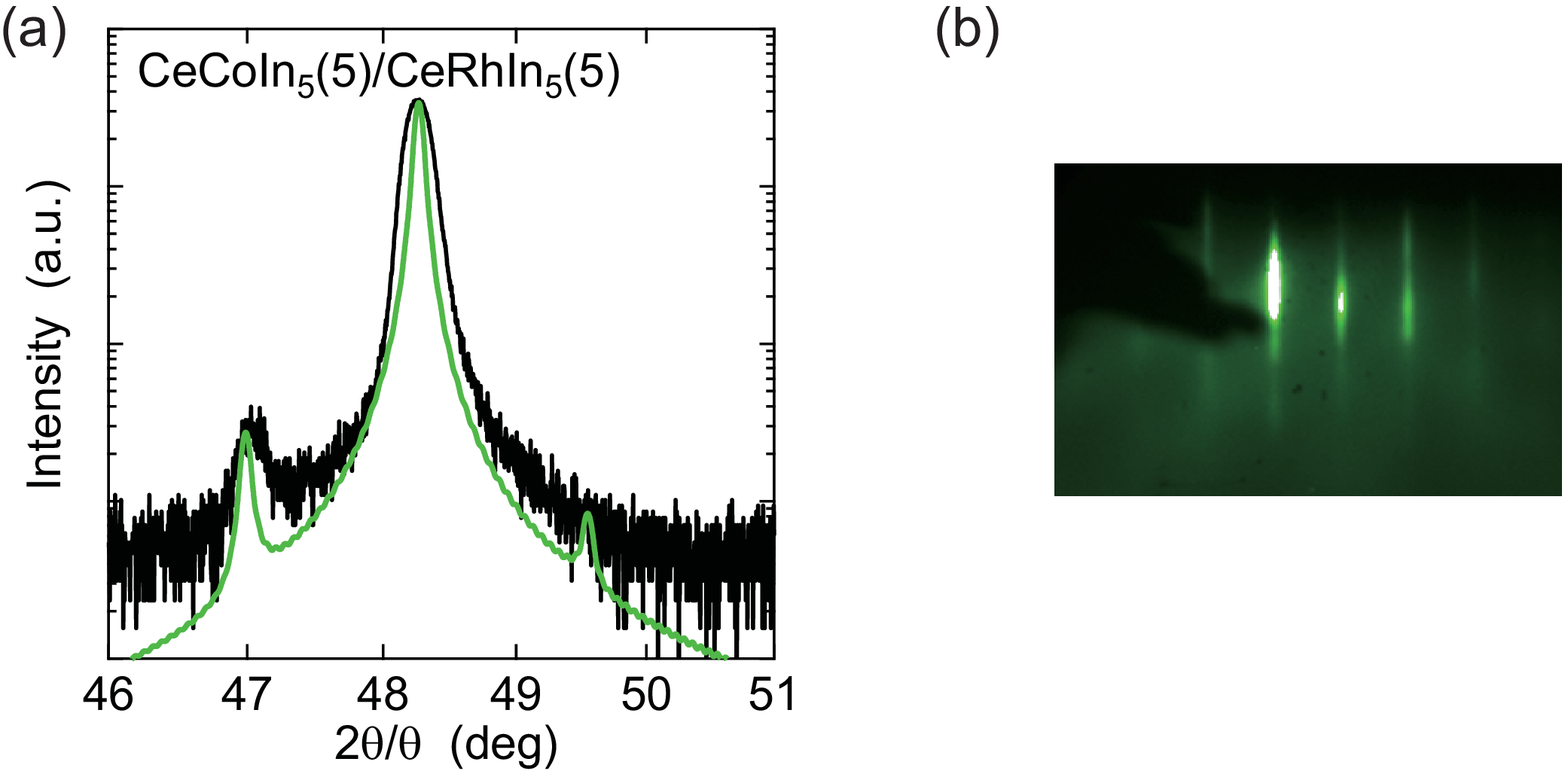}
	\caption{(a) Cu $K\alpha_1$ X-ray diffraction pattern for CeCoIn$_5$(5)/CeRhIn$_5$(5) superlattice with total thickness of 300\,nm.  In addition to the [0,0,4] main peak at $2\theta\sim48^{\circ}$,  satellite peaks are observed.  The positions of the satellite peaks and their asymmetric heights can be reproduced by the step-model simulations (green line) ignoring interface and layer-thickness fluctuations.  (b)  The reflection high-energy electron diffraction (RHEED) image taken after the crystal growth.   The streak patterns in the RHEED image indicate the epitaxial growth of each layer with atomic flatness. The streak patterns are observed during the crystal growth. }
\end{figure}

\newpage

\begin{figure}[t]
	\includegraphics[width=\linewidth]{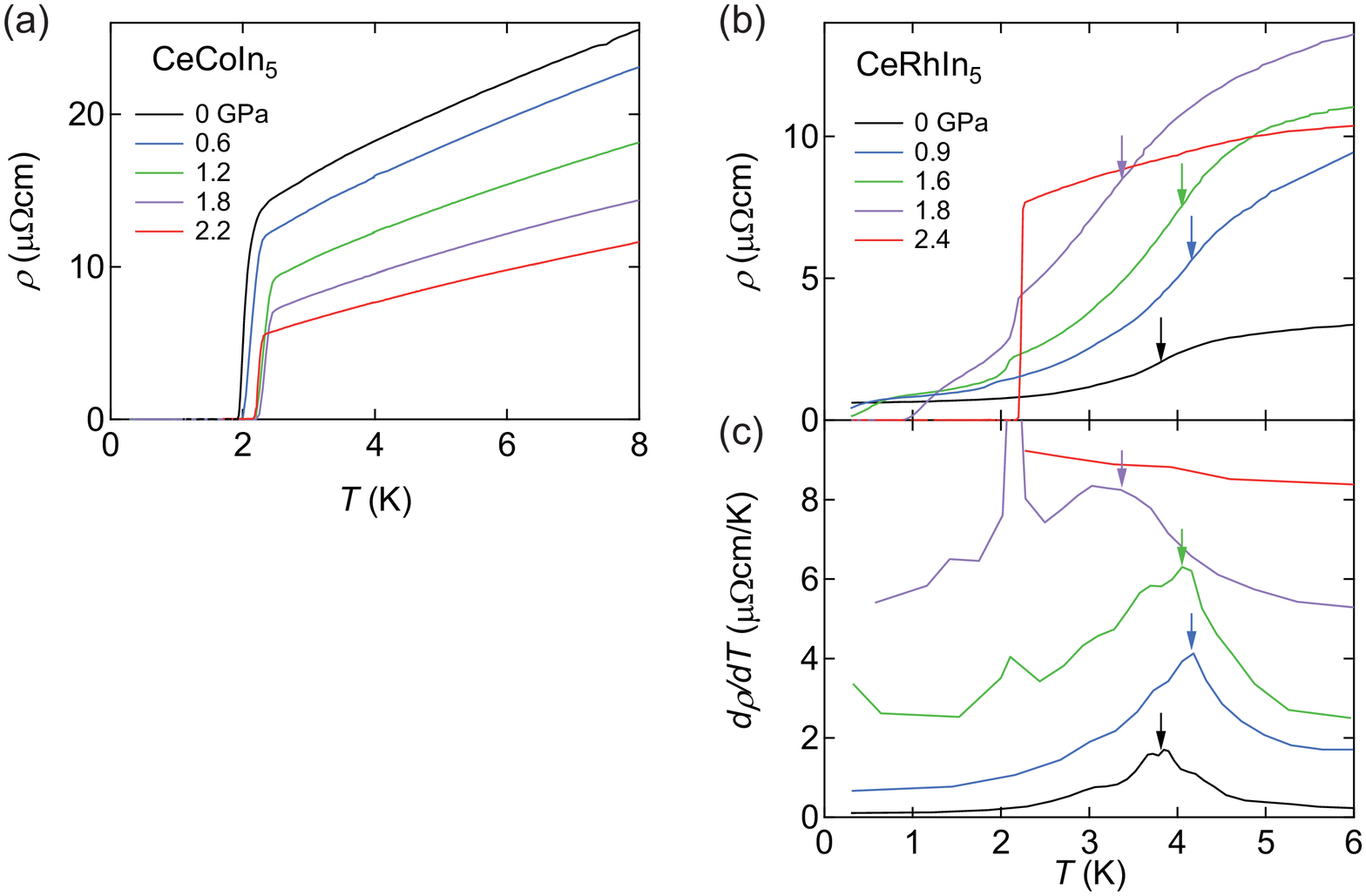}
	\caption{(a) Temperature dependence of the resistivity of CeCoIn$_5$ thin films with thickness of 300\,nm under pressure.  (b) The resistivity of CeRhIn$_5$ thin films with thickness of 300\,nm under pressure.  (c) Temperature derivative of the resistivity.  }
\end{figure}

\newpage

\begin{figure}[t]
	\includegraphics[width=0.6\linewidth]{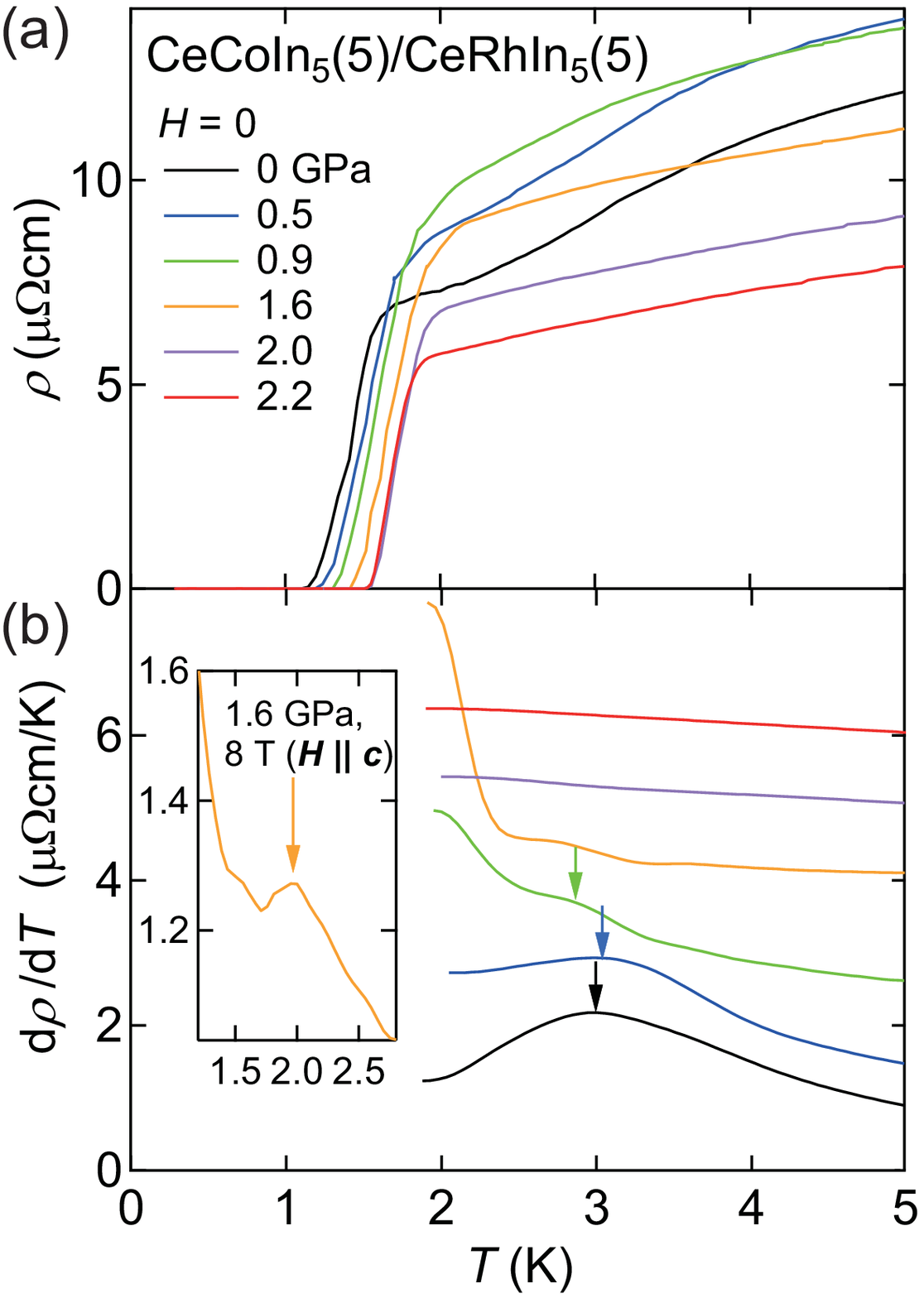}
	\caption{(a) Temperature dependence of the resistivity of CeCoIn$_5$(5)/CeRhIn$_5(5)$ superlattice under pressure. (b) Temperature derivative of the resistivity $d\rho/dT$. The inset shows $d\rho/dT$ at 1.6\,GPa  in perpendicular field of 8\,T.  The AFM transition temperature is determined by the peak of $d\rho/dT$.}
\end{figure}

\newpage

\begin{figure}[h]
	\includegraphics[width=1.0\linewidth]{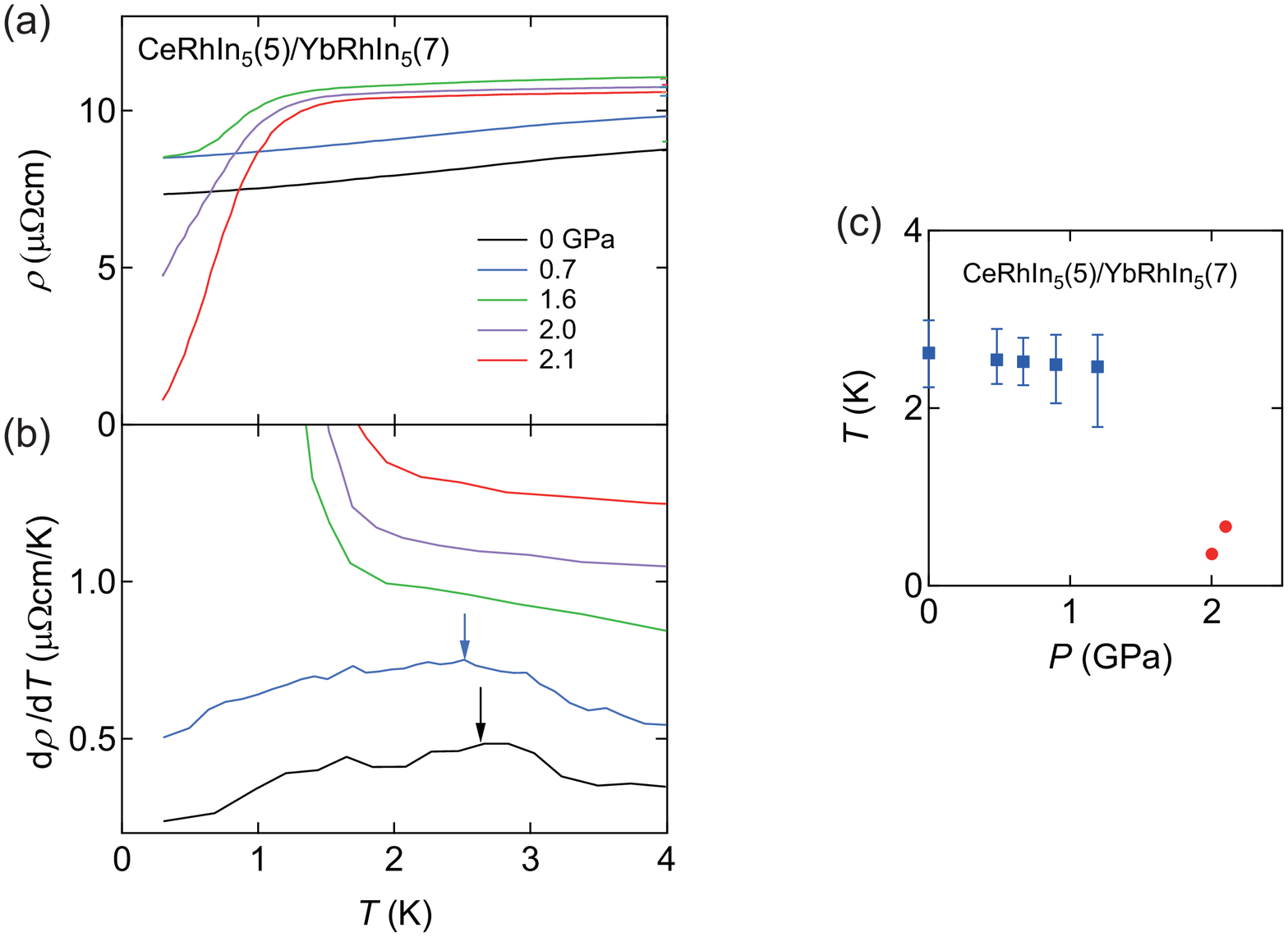}
	\caption{(a) Temperature dependence of the resistivity of  CeRhIn$_5$(5)/YbRhIn$_5$(7) superlattice under pressure. (b) Temperature derivative of the resistivity $d\rho/dT$. The AFM transition temperature is determined by the peak of $d\rho/dT$. (c) $p$-$T$ phase diagram determined by the resistivity. Blue squares and red circles represent the AFM and superconducting transition temperatures, respectively.   }
\end{figure}

\newpage

\begin{figure}[h]
	\includegraphics[width=0.8\linewidth]{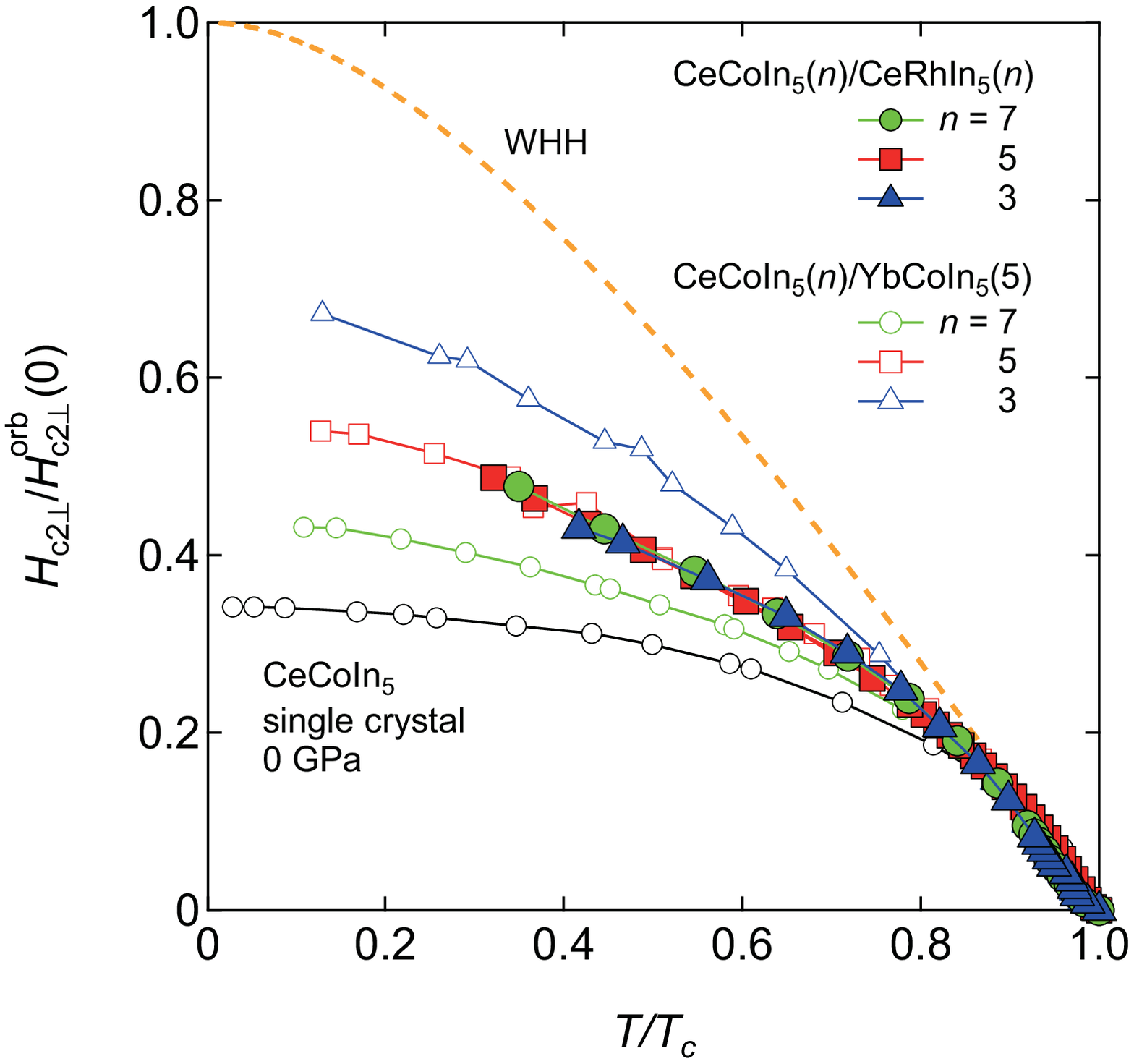}
	\caption{Out-of-plane upper critical field $H_{c2\perp}$ normalized by the orbital-limited upper critical field at $T=0$\,K, $H_{c2\perp}/H_{c2\perp}^{\rm orb}(0)$, for CeCoIn$_5$($n$)/CeRhIn$_5$($n$) and CeCoIn$_5$($n$)/YbCoIn$_5$(5) superlattices with $n=7$, 5, and 3  is plotted as a function of the normalized temperature $T/T_c$.   Two extreme cases, i.e., the result of  the bulk CeCoIn$_5$ dominated by Pauli paramagnetic effect and the WHH curve with no Pauli effect, are also shown.  }
\end{figure}

\end{document}